# Designing Recommender Systems to Depolarize

**Jonathan Stray**
Center for Human-Compatible AI
University of California at Berkeley
jstray@berkeley.edu

**Abstract**

Polarization is implicated in the erosion of democracy and the progression to violence, which makes the polarization properties of large algorithmic content selection systems (recommender systems) a matter of concern for peace and security. While algorithm-driven social media does not seem to be a primary driver of polarization at the country level, it could be a useful intervention point in polarized societies. This paper examines algorithmic depolarization interventions with the goal of conflict transformation: not suppressing or eliminating conflict but moving towards more constructive conflict. Algorithmic intervention is considered at three stages: which content is available (moderation), how content is selected and personalized (ranking), and content presentation and controls (user interface). Empirical studies of online conflict suggest that the exposure diversity intervention proposed as an antidote to "filter bubbles" can be improved and can even worsen polarization under some conditions. Using civility metrics in conjunction with diversity in content selection may be more effective. However, diversity-based interventions have not been tested at scale and may not work in the diverse and dynamic contexts of real platforms. Instead, intervening in platform polarization dynamics will likely require continuous monitoring of polarization metrics, such as the widely used "feeling thermometer." These metrics can be used to evaluate product features, and potentially engineered as algorithmic objectives. It may further prove necessary to include polarization measures in the objective functions of recommender algorithms to prevent optimization processes from creating conflict as a side effect.

## 1   Introduction

Polarization is a condition where myriad differences in society fuse and harden into a single axis of identity and conflict (Iyengar & Westwood, 2015), and has been increasing for multiple decades in several democracies (Boxell et al., 2020; Draca & Schwarz, 2018). Comparative studies that examine polarization across countries argue that increasing polarization is a contributing factor to the democratic erosion seen in many countries, including Venezuela, Hungary, Turkey, and the United States (McCoy et al., 2018; Somer & McCoy, 2019). Polarization produces a feedback loop where diverging identities lead to less intergroup contact which in turn leads to increased polarization, culminating in a hardened us-vs-them mentality that can contribute to the deterioration of democratic norms. Most conflict escalation models consider polarization a key part of the feedback dynamics that lead to violent conflict (Collins, 2012). Peace and security demand that we address situations of increasing polarization, which is why the international peacebuilding community concerns itself with polarization (Ramsbotham et al., 2016).

Scholars have long studied the relation between media and conflict, a tradition that now includes digital media (Hofstetter, 2021; Tellidis & Kappler, 2016) much of which is algorithmically selected and personalized. The algorithms that choose which items are shown to each user are called recommender systems and all major news aggregators and social media platforms have such a system at their core. Modern recommender systems select content based on a variety of information sources such as the content of each item, a user's expressed preferences, their past consumption behavior, the behavior of similar users, user survey responses, fairness considerations, and more (Aggarwal, 2016). Note that "recommender" is a computer science term of art that covers all algorithmic content selection on the basis of implicit information, i.e. not as the result of a search query. This content might be presented as "recommended for you," labelled as "news" or "trends," or appear as a feed or timeline.

There has been intense interest in the question of whether recommender systems affect large-scale conflict dynamics. Most of the work on recommenders and polarization has taken place within the "filter bubble" paradigm and therefore explored the idea of exposure diversity (Helberger et al., 2018). Selective exposure is the idea that individuals will preferentially choose news sources and articles that are ideologically aligned (Prior, 2013). Because recommender systems respond to user interests, there is the possibility of a feedback loop where both recommendations and user interests progressively narrow. Indeed, simulations have demonstrated such polarization-increasing effects in stylized settings (Jiang et al., 2019; Krueger et al., 2020; Rychwalska & Roszczyńska-Kurasińska, 2018; Stoica & Chaintreau, 2019).

However, available evidence mostly disfavors the hypothesis that recommender systems are driving polarization through selective exposure, aka "filter bubbles" or "echo chambers" (Bruns, 2019; Zuiderveen Borgesius et al., 2016). Algorithmically personalized news seems to be quite similar for all users (Guess et al., 2018), is typically no less diverse than selections by human editors (Möller et al., 2018), and social media users consume a more diverse range of news sources than non-users (Fletcher & Nielsen, 2018). Most recently, Feezell et al. (2021) find no difference in affective polarization scores between Americans who get their news from conventional sources vs. social media.

Non-news personalized content could still be polarizing. Lelkes et. al. (2017) compare the introduction of broadband access across U.S. states from 2004-2008 and find a small causal increase in affective polarization. Yet polarization began increasing in the U.S. decades before social media, and is increasing faster among individuals aged 65 and up, a demographic with low internet usage (Boxell et al., 2017). A cross-country analysis shows no clear relationship between polarization and increasing internet usage, as many OECD countries with high internet usage such as Britain, Sweden, Norway and Germany show decreasing affective polarization (Boxell et al., 2020).

Direct experimental intervention is probably the best way to study the causality of recommender systems. Allcott et. al. (2020) paid U.S. users to stay off Facebook for a month and found that an index of polarization measures decreased by 0.16 SD (standard deviations). This may have been due to a decrease in exposure to polarizing posts, comments, and discussions, but this intervention also decreased time spent on news by 15 percent, and news consumption can itself be polarizing (Martin & Yurukoglu, 2016; Melki & Sekeris, 2019). By contrast, in a similar study in Bosnia and Herzegovina users who deactivated Facebook during a genocide remembrance week showed *greater* polarization, a 0.24 SD increase on an index of ethnic polarization (Asimovic et al., 2021). The increase was smaller for users who had a more ethnically diverse offline social group, suggesting that Facebook was in this case providing depolarizing diversity. While these studies suggest causation, the effects are not unidirectional or straightforward.

Rather than asking if social media is driving polarization, it may be more productive to ask if social media interventions can decrease polarization. The main contribution of this paper is to propose several methods for building recommender systems that actively reduce polarization.

Note that polarization is conceptually distinct from radicalization. Polarization is a process that "defines other groups in the social and political arena as allies or opponents" while radicalization involves people who "become separated from the mainstream norms and values of their society" and may engage in violence (van Stekelenburg, 2014). There is a growing body of work studying the connection between recommender systems and radicalization (Baugut & Neumann, 2020; Hosseinmardi et al., 2020; Ledwich & Zaitsev, 2019; Munger & Phillips, 2020; Ribeiro et al., 2020) but this is methodologically challenging, and has not yet established a robust causal link. While social media is plausibly involved in radicalization processes the nature of this connection is complex and poorly understood. This work concerns polarization only, arguing that polarization itself is a bad outcome and a precursor to more extreme conflict.

In this paper I first make the moral argument for attempting to reduce polarization through recommender systems, framing it as a conflict transformation intervention. I then review definitions and metrics of polarization before considering depolarization interventions at three stages: which content is available (moderation), how content is selected and personalized (ranking), and content presentation and controls (user interface). The most commonly proposed depolarization intervention is exposure to ideologically diverse content, but this may not be effective because mere exposure does not necessarily depolarize, and sometimes polarizes further. While there are other promising approaches such as exposure to *civil* counter-ideological content, these may not be sufficiently robust to the incredibly diverse conditions of real-world platforms. Instead, I propose continuously monitoring survey measures of affective polarization so as to drive recommender outcomes in a feedback loop. Polarization metrics can be used both at the managerial level and at the algorithmic level, potentially through reinforcement learning.

## 2  Depolarization as conflict transformation

There are complicated questions around intervening in societal conflicts through media, and additional concerns around the use of AI for this purpose. At worst, algorithmically suppressing disagreements could amount to authoritarian pacification. The Chinese social media censorship regime is an instructive example of democratically questionable interventions in the name of harmony (Creemers, 2017; G. King et al., 2017). Therefore, I frame the goal of depolarization as conflict transformation: not eliminating or resolving conflict but making that conflict better in some way, e.g. less prone to violence and more likely to lead to justice (Jeong, 2019).

Indeed, it's not clear that platform users *want* to be "depolarized," and in any mass conflict situation there will be people who argue for escalation in the strongest moral terms. There is a corresponding line of argument that polarization is beneficial. Political theorists have argued that polarization reduces corruption by increasing accountability (Melki & Pickering, 2020) and generally helps differentiate political parties in a way that provides a meaningful choice to voters. In the mid 20$^{th}$ century mainstream political scientists worried that America wasn't polarized *enough* (American Political Science Association, 1950). Importantly, fights for justice or accountability can also increase polarization, such as the American Civil Rights movement of the 1960s (D. S. King & Smith, 2008). There are parallels to the idea of a just war.

Yet polarization also has severe downsides. Polarization at the elite level causes "gridlock" that makes effective governance difficult (F. E. Lee, 2015) but contemporary polarization reaches far beyond lawmakers. The politicization of all spheres of society destroys social bonds at the family, community, and national levels (A. H. Y. Lee, 2020). By some measures, cross-partisan dislike in the U.S. is now considerably stronger than racial resentment, and has large effects on social choices such as hiring, university admissions, dating, family relations, friendships, and purchasing decisions (Iyengar et al., 2018). Polarization erodes the norms that constrain conflict escalation, leading to "morally outrageous" behavior on all sides (Deutsch, 1969), and is a key precursor to violence (Collins, 2012). Ultimately, polarization appears to be a causal factor in the destruction of democracies (McCoy et al., 2018; Somer & McCoy, 2019).

There is a tension between peace and justice. Actions that promote peace may make justice harder, and vice versa. Yet a democracy requires both, an observation which leads to the concept of a just peace (Fixdal, 2012). Instead of trying to eliminate conflict, we can try to understand what makes it good or bad. In an agonistic theory of democracy it is considered normal for political adversaries to be engaged in "opposing hegemonic projects," and conflict is not to be eliminated but "tamed" (Mouffe, 2002). Perhaps the most sophisticated understandings of conflict come from the peacebuilding tradition, which came into its own as an applied discipline after World War II. Fifty years ago, Deutsch described the difference between "constructive" and "destructive" conflict, with particular attention to the dynamics of escalation:

> Paralleling the expansion of the scope of conflict there is an increasing reliance upon a strategy of power and upon the tactics of threat, coercion, and deception. Correspondingly, there is a shift away from a strategy of persuasion and from the tactics of conciliation, minimizing differences, and enhancing mutual understanding and good-will. And within each of the conflicting parties, there is increasing pressure for uniformity of opinion and a tendency for leadership and control to be taken away from those elements that are more conciliatory and invested in those who are militantly organized for waging conflict through combat.
> …
> It leads to a suspicious, hostile attitude which increases the sensitivity to differences and threats, while minimizing the awareness of similarities. This, in turn, makes the usually accepted norms of conduct and morality which govern one's behavior toward others who are similar to oneself less applicable. Hence, it permits behavior toward the other which would be considered outrageous if directed toward someone like oneself. (Deutsch, 1969)

On the other hand, Lederach describes how conflict is necessary for positive social change and how conflict transformation moves towards better conflict processes:

> A transformational approach recognizes that conflict is a normal and continuous dynamic within human relationships. Moreover, conflict brings with it the potential for constructive change. Positive change does not always happen, of course. As we all know too well, many times conflict results in long-standing cycles of hurt and destruction. But the key to transformation is a proactive bias toward seeing conflict as a potential catalyst for growth.
> …
> A transformational approach seeks to understand the particular episode of conflict not in isolation, but as embedded in the greater pattern. Change is understood both at the level of immediate presenting issues and that of broader patterns and issues.
> (Lederach, 2014)

Or as Ripley puts it:

> The challenge of our time is to mobilize great masses of people to make change without dehumanizing one another. Not just because it's morally right but because it works. (Ripley, 2021, p. 13)

Polarization is potentially an important intervention point in conflict dynamics because it is involved in escalation pathways through multiple routes. Polarization can be exploited for political mobilization through us-versus-them rhetoric, as has long been understood by activists (Layman et al., 2010) and other "political entrepreneurs" (Somer & McCoy, 2019) and as demonstrated by the fact that the most politically engaged citizens are found at the ideological extremes (Pew Research Center, 2014). However, this kind of exploitation further increases polarization. Indeed, polarization is involved in a variety of pernicious feedback loops: polarization leads to less intergroup contact, which causes polarization (A. H. Y. Lee, 2020); polarization is a precursor to violence, which causes polarization (Collins, 2012); polarization leads to selective information exposure, which causes polarization (Kim, 2015) and so on. These causal dynamics suggest that polarization could be an important intervention point in conflict escalation.

Conflicts that involve democratic erosion or violence are deeply troubling, to the point where conflict-transforming interventions may be warranted on human rights grounds. In the U.S. support for violence in service of political ends is increasing on both the left and the right (Diamond et al., 2020). In short, partisans are willing to violate democratic norms when polarization is high. A recent review concluded that "the goal of these [depolarizing] interventions is to move toward a system in which the public forcefully debates political ideals and policies while resisting tendencies that undermine democracy and human rights." (Finkel et al., 2020)

# 3 Measuring polarization

Quantitative measures are needed to evaluate polarization at scale. This is not merely a problem of measurement, but of definition. Polarization has been studied through differences in legislative voting patterns (Hare & Poole, 2014) and the language used in U.S. Congressional speech (Gentzkow et al., 2017). At the population level it has been operationalized as the increasing correlation of policy preferences over multiple issues (Draca & Schwarz, 2018; Kiley, 2017) and as increasing animosity towards the political outgroup, known as affective polarization (Iyengar & Westwood, 2015). All of these indicators show increasing polarization in the US over the last 40 years. Globally the results are more mixed, with some OECD countries experiencing increasing polarization and others showing flat or decreasing trends (Boxell et al., 2020; Draca & Schwarz, 2018).

Affective polarization has become a key concept in the analysis of American politics as "ordinary Americans increasingly dislike and distrust those from the other party" (Iyengar et al., 2018). Affective polarization is a consequence of partisan identity, which is a better model of contemporary political conflict than differences in issue positions (Finkel et al., 2020). It also has the advantage of being operationalizable through straightforward survey measures, such as the *feeling thermometer* which is one of the oldest and most widely used polarization measures. This method asks respondents to rate their feeling about each political party on a scale from 0 (cold) to 100 (warm). The difference in scores, the *net feeling thermometer*, is taken to be a measure of affective polarization. This question has been asked on the American National Election Study since the 1970s, and is frequently used in studies of polarization and social media (Feezell et al., 2021; Levy, 2020; Suhay et al., 2018). While there are different measures of affective polarization they are mostly highly correlated (Druckman & Levendusky, 2019).

Affective polarization – negative feelings about the "other side" – has serious interpersonal consequences. Tellingly, 13 percent of Americans reported that they had ended a relationship with a family member or close friend after the 2016 election (Whitesides, 2017). Affective polarization correlates with dehumanization, "a significant step toward depriving individuals who belong to certain groups or categories of individual-level depth or complexity of feelings, motivation, or personality" (Martherus et al., 2021). It leads to the destruction of social bonds and increased outgroup prejudice across all facets of social and political life (Iyengar et al., 2018; A. H. Y. Lee, 2020; Somer & McCoy, 2019). In short, affective polarization now strongly colors the experience of daily life and relationships in multiple countries and has potentially grim consequences for democracy.

# 4 Algorithmic depolarization interventions

Recommender-based systems such as social media and news aggregators are more than just "algorithms," and an analysis of the polarization effects of this wide array of products and platforms could potentially be very broad. To narrow the scope, I will consider three key places where changes to recommender systems might be used for depolarization:

**Which content is available (moderation).** Much previous work on polarization has concerned itself with which content is allowed on a platform. For example, hate speech and incitements to violence are routinely removed through a combination of human moderators, machine learning classifiers, and user flagging.

**How content is selected (ranking).** Algorithmic content selection is essentially a prioritization problem, and all contemporary recommendation systems score each item based on a number of criteria. An intervention in content ranking addresses the core question of who sees what. Most of the approaches considered in this paper are modifications to content ranking.

**How content is presented (interface).** Selected items are presented to the user in some way, who can interact with the recommender system through predefined controls. Different presentations or different controls may be conducive to better or worse conflict.

It should immediately be said that there are many possible non-algorithmic social media depolarization interventions, such as community moderation (Jhaver et al., 2017). There are also hybrid approaches, like The Commons which uses automated messages (social media bots) to find people who want to engage in depolarizing conversations, then connects them to human facilitators (Build Up, 2019). There are also a wide variety of depolarization strategies entirely outside of algorithmic media, such as approaches based in journalism, politics, or education, any of which may prove to be more effective. Nonetheless this paper considers only algorithmic interventions in recommender systems because algorithmic content selection has been a central topic of concern, automation provides a path to scaling interventions, and the polarization properties of recommender algorithms are important in any case.

### 4.1 Removing polarizing content

Many kinds of content are now removed from platforms, including spam, misinformation, hate speech, sexual material, criminal activity, and so on (Halevy et al., 2020). While the removal of violent material and incitements to violence may be particularly important in the context of an active conflict (Schirch, 2020), the removal of less extreme material is a blunt approach that may not be justified as a mass depolarization intervention.

This kind of content removal is often called "moderation," but it's important to distinguish between community moderation and algorithm-assisted moderation at scale. At the level of an online community or discussion group, volunteer moderators are able to set and enforce norms that lead to productive discussion of polarized topics, as a study of the r/ChangeMyView subreddit shows (Jhaver et al., 2017). Such studies of the micro-dynamics of conflict provide important clues for potential depolarization interventions. Moderators remove posts and suspend accounts, but they also state reasons for their actions, take part in discussions about appropriate policy, and consider appeals.

Platform moderation, by contrast, operates at vast scale to identify unwanted content through a combination of paid moderators and machine learning models. It is acontextual, impersonal, and difficult to appeal (York & Zuckerman, 2019). The low rates of offending content mean that true positives (correctly removed material) may be vastly outnumbered by false positives (incorrectly removed material) unless automated classifiers can be made unrealistically accurate (Duarte et al., 2017). Further, content removal is concerning from a freedom of expression perspective, and the standards for removal are widely contested (Keller, 2018). Facebook alone is "most certainly the world's largest censorship body" (York & Zuckerman, 2019).

Given these concerns, there should be a high bar for automated content removal as a mass depolarization intervention. What should be the standard for unacceptably polarizing material? We could algorithmically remove all angry political comments, but do we want to? Removing all material which might intensify conflict would leave the public sphere arid, authoritarian and devoid of any real politics.

### 4.2 Increasing exposure diversity

Most prior work on the relationship between polarization and social media has been based on the concept of exposure diversity. The most frequently proposed fix is to algorithmically increase the diversity of social media users' feeds (Bozdag & van den Hoven, 2015; Celis et al., 2019; Helberger et al., 2018) and a variety of recommender diversification algorithms have been developed (Castells et al., 2015). This is intuitively appealing, as inter-group contact has been demonstrated to reduce prejudice (Pettigrew & Tropp, 2006).

This approach presupposes that a lack of diversity in online media content is causing polarization, which is questionable as discussed above. "Diversity" is also poorly defined, and may refer to source diversity, topic diversity, author diversity, audience diversity, and more. A review of media diversity by Loecherbach et. al. (2020) notes that "research on this topic has been held back by the lack of conceptual clarity about media diversity and by a slow adoption of methods to measure and analyze it." Further, the causal connection between exposure diversity and polarization is complex and under some conditions exposure to outgroup content can actually increase polarization (Bail et al., 2018; Paolini et al., 2010; Rychwalska & Roszczyńska-Kurasińska, 2018; Taber & Lodge, 2006).

Yet increasing exposure diversity can work, at least somewhat. One experiment tested the effect of asking US Facebook users to subscribe to ("like") up to four liberal or conservative news outlets, measuring changes in affective polarization through a survey two weeks later. This level of exposure to outgroup information decreased affective polarization by about 1 point on a 100-point scale (Levy, 2020). By comparison, the rate of increase in affective polarization in the U.S. since 1975 is estimated at 0.6 points per year (Finkel et al., 2020). Rescaled to the same 100 point scale, the previously discussed experiment of leaving Facebook for a month resulted in about a 2 point decrease (Allcott et al., 2020, p. 652) though only on issue-based rather than affective measures. All of these estimates should be considered quite rough.

This demonstrates that increased exposure diversity can be a useful intervention point for depolarization, but the effect so far has been modest. Are different or better approaches possible? For example, Levy (2020) tested only *news* diversity, meaning professional journalism. Polarization may turn out to be more sensitive to non-news content or user comments.

### 4.3 Recommending civil arguments

Several studies have attempted to determine the conditions under which polarization and depolarization occur. Kim & Kim (2019) found that those who read uncivil comments arguing for an opposing view rated themselves as closer to ideological extremes on a post-exposure survey than those who did not. Civility may not be depolarizing per se, but incivility does seem to be polarizing. Suhay et al. (2018) similarly show that comments that negatively describe political identities (e.g. "Liberals are ignorant") increase polarization as measured by the feeling thermometer question. This effect also appears in the context of partisan media sources (e.g. MSNBC, Fox) where "incivility [of] out-party sources affectively polarizes the audience" (Druckman et al., 2019).

It seems likely that "civility" and "partisan criticism" can be algorithmically scored through existing natural language processing techniques, drawing on previous work classifying hate speech and harassment. All are conceptually close to the "toxicity" operationalized by contemporary comment classification models (Noever, 2018). While these models are mostly used for moderation -- that is, removing offending comments – they could also provide a "civility" signal that is incorporated into recommender item ranking. Twitter has experimented with this idea (Wagner, 2019) but I am not aware of any production recommender that incorporates a civility signal in content ranking (as opposed to content moderation).

In addition to demoting uncivil content, it is possible to promote civil content. Experimental evidence shows that ranking high-quality comments at the top can positively alter the tone of subsequent discussion (Berry & Taylor, 2017). In effect, this intervention hopes to model respectful disagreement. This may not work if there are not many natural examples of productive inter-group conversation. In particular, there may be a lack of journalism content that takes a depolarizing approach to reporting on controversial issues (Hautakangas & Ahva, 2018; Prior & Stroud, 2015; Ripley, 2018).

Of course, uncivil language can be necessary and important. We certainly don't want an algorithmic media system that redirects attention away from anyone raising their voice. Indeed, several theories of democracy require such confrontation, such as critical approaches (Helberger, 2019) or agonistic models (Mouffe, 2002). Hence, there is a tension between encouraging expression and intervening to make the conversation more productive – this is the art of (algorithmic) mediation.

### 4.4 Priming better interactions

Given a particular set of items selected for a user, it may be possible to present them in a way that encourages more productive conflict. Language seems particularly important in political disagreements. Intriguingly, replacing the usual "like" button with a "respect" button increased the number of clicks on counter-ideological comments, that is, people were more likely to "respect" something they disagreed with than to "like" it (Stroud et al., 2017).

While civility norms have been shown to contribute to successful online discussions of polarized topics (Jhaver et al., 2017) it is difficult to automate the promulgation and enforcement of such norms. One intriguing possibility is to change the content of automated messages, such as the message welcoming someone to a group. In a large scale experiment on r/science on Reddit, adding a short note explaining what types of posts will be removed and noting that "our 1200 moderators encourage respectful discussion" greatly reduced the rate at which newcomers violated community norms (Matias, 2019).

In a sense, changing user behavior is the strongest depolarization intervention. This is not at all simple to accomplish, but these studies demonstrate that simple user interface changes can have profound effects.

## 5   Learning to depolarize

The approaches discussed above are justified on the basis of sociological theory, from results in laboratory settings, or through modest platform experiments. Real platforms are enormous, diverse, and dynamic environments, and ecological validity is a serious problem for the development of social media interventions (Griffioen et al., 2020). It is likely to be difficult to predict which depolarization interventions will succeed. The best approach will vary between subgroups, in different contexts, and over time.

Effective management of polarization will therefore depend on continual monitoring of polarization outcomes by platform operators. Affective polarization measures may prove to be the most useful category of metrics, in part because they are agnostic to the type of content that drives polarization. More cognitive measures of polarization, such as issue position surveys (Draca & Schwarz, 2018; Kiley, 2017) may be less diagnostic for social media, where many interactions will not involve discussions of substantial policy preferences.

Platforms already monitor various non-engagement measures and incorporate them into recommender design and ranking (Stray, 2020). Facebook asks users whether specific posts led to a *meaningful social interaction* on or off the platform. This is a construct from social psychology that appears to be similarly interpretable across cultures (Litt et al., 2020). YouTube similarly incorporates user satisfaction ratings obtained by asking users what they thought of specific recommendations (Zhao et al., 2019). Such metrics are used to drive product choices at the managerial level by selectively deploying changes, a form of A/B testing. They are also incorporated directly into the predictive models underlying item ranking, as the next section describes, but the first and most fundamental depolarization intervention is simply to monitor for actual polarization outcomes, rather than betting on theory.

### 5.1   Optimizing for depolarization

Survey responses can be used to train recommender ranking algorithms, for example by building a model that predicts whether an item is going to lead to a positive survey answer for a particular user in a particular context. This is, technically speaking, similar to predicting which items will result in a click. Optimizing for predicted survey responses is an important technique in the nascent field of recommender alignment, the practice of getting recommender systems to enact human values (Stray, 2021; Stray et al., 2020).

The feeling thermometer has been used experimentally to evaluate the polarizing effect of seeing a post, by taking the difference between treatment and control groups (Kim & Kim, 2019; Suhay et al., 2018). If it proves possible to know whether individual posts or conversations are polarizing, it should be possible to build a model to predict the polarization effect of showing novel posts. Similar classifiers are already in use to detect misinformation, hate speech, bullying etc. One plausible technique is the TIES model, which takes into account not only the text and image content of a specific post but the sequence of interactions around it, including discussions in comments, likes, shares, etc. (Noorshams et al., 2020). In the context of an online discussion, the goal would be to determine whether users are having a productive exchange of views or a divisive argument, so the history of interactions carries significant information.

Alternatively, affective polarization measures could be used longitudinally, perhaps by asking a panel of users to respond to a feeling thermometer question daily or weekly, thereby measuring attitudes over time. When compared to a control group, this amounts to a difference-in-differences design which gives robust causal estimates under certain assumptions (Angrist & Pischke, 2009, Chapter 5). That is, it should be possible to learn the actual polarizing effects of selecting different distributions of items. However, using longitudinal data to drive recommendation systems toward selecting depolarizing content is technically challenging due to the much longer time scale and higher level of abstraction as compared to feedback on individual items.

Reinforcement learning (RL) algorithms may be the most general and powerful approach to learning patterns of recommendation which optimize long term outcomes (Ie et al., 2019; Mladenov et al., 2019). In principle, affective polarization survey measures could be used as a reward signal for reinforcement learning-based recommenders. However, this sort of learning from sparse survey feedback has not yet been demonstrated. Additional algorithmic development will be necessary before longitudinal polarization measures can be incorporated into content selection algorithms, but the necessary technical research is underway because other sparse, long term signals such as user subscriptions have immediate business value.

In other words, the same methods that make it possible to predict what movies to show someone to get them to subscribe may also make it possible to learn which patterns of interaction increase or reduce polarization.

### 5.2 Unintended consequences and the necessity of specification

The effective use of sociological metrics is complicated and can fail in a number of ways, regardless of whether the metric is used by people or algorithms. Using reinforcement learning to attempt large scale political intervention should be a particularly alarming prospect. While there is a strong moral case for designing recommender systems to depolarize, unintended consequences could swamp any positive effects.

A metric is an operationalization of some theoretical construct, and might be an invalid measure for a variety of reasons (Jacobs & Wallach, 2019). Even a well-constructed metric almost never represents what we really care about: clickbait lies entirely in the difference between "click" and "value." When used as targets, metrics suffer from a number of problems involving gaming and spurious correlations, which can be understood in causal terms as variations of Goodhart's law (Manheim & Garrabrant, 2018). It is particularly important to undertake ongoing qualitative methods and user research, to know whether current metrics are adequately tracking the intended goals – and to learn of whatever else may be happening.

Metrics often fail when used in management contexts because they are irrelevant, illegitimate, gamed, or aren't updated as the context changes (Jackson, 2005). Using metrics to train a powerful optimizing system introduces further concerns (Thomas & Uminsky, 2020). Different effects for different subgroups may be a particular problem for recommender systems which typically optimize average scores (Li et al., 2021). While it's always useful to monitor for slippage between a metric's intent and what it is actually measuring, this is particularly important when a measure becomes the target of society-wide AI optimization (Stray, 2020). If we choose to apply reinforcement learning to polarization metrics, those metrics will require continuous evaluation.

On the other hand, *not* using polarization measures in algorithmic content selection may be far worse. Optimization algorithms which do not penalize polarization measures might learn, as humans do, that polarization can be exploited for engagement. Or they might merely increase conflict as an agnostic side effect, which is no better. In general, under-specification is a serious hazard in the creation of machine learning models (D'Amour et al., 2020). If we do not specify the intended effect of a recommender system on polarization, we should not be surprised to find unexpected outcomes.

### 6 Conclusion

Polarization is a hardening division of society into "us" vs "them." It interacts with a number of conflict feedback processes and eventually leads to democratic erosion and violence (McCoy et al., 2018; Somer & McCoy, 2019). The goal of a depolarization intervention is not to suppress conflict, but to have better conflict that moves towards constructive societal change (Deutsch, 1969; Jeong, 2019; Lederach, 2014; Ripley, 2021). While all societies face complex tensions between peace and justice, depolarization interventions may ultimately be justified on human rights grounds (Finkel et al., 2020) just as other peacebuilding interventions are.

Available evidence suggests that social media usage is not driving increases in polarization at the country level (Boxell et al., 2017, 2020). In particular, there is little empirical support for the idea that personalization is reducing exposure to diverse information (Guess et al., 2018; Zuiderveen Borgesius et al., 2016). Nonetheless, there is some evidence that social media-based interventions can reduce polarization among users. A recent experimental test of increasing news diversity produced a small decrease in polarization (Levy, 2020). Paying users to stay off Facebook for a month produced small decreases in issue polarization, though not affective polarization (Allcott et al., 2020).

Moderation, the removal of unwanted content, can be important especially in the context of a violent conflict (Schirch, 2020) but it is probably too blunt an instrument for depolarization. Content ranking defines what each user sees and is the most general intervention point. While exposure to diverse perspectives can actually increase polarization (Bail et al., 2018) increased exposure diversity does depolarize in some contexts (Levy, 2020; Pettigrew & Tropp, 2006). Recommenders could augment diversity by de-prioritizing content that has been shown to be polarizing, including uncivil presentations of outgroup opinions (Kim & Kim, 2019) and criticism of partisan identities (Suhay et al., 2018). Content presentation and user interface may also have depolarization effects, as has been shown in experiments changing "like" to "respect" (Stroud et al., 2017) and adding a message reminding users of community norms (Matias, 2019).

Yet none of the above approaches directly target the outcome of interest. Any depolarization method based on selecting content according to pre-existing theory may prove unable to cope with the radically diverse and dynamic contexts of a real recommender system. The solution is to directly and continuously measure polarization outcomes.

Existing polarization measures, particularly affective polarization measures, have been used to evaluate the effect of encountering different types of comments on news articles (Kim & Kim, 2019) and the same methods should generalize to other types of items including user posts, discussion threads, and so on. Such survey data can be used to evaluate recommender system changes and make deployment decisions. It can also be used to train polarization prediction models, much as existing recommender models predict meaningful social interactions and other survey results (Stray, 2020). Ultimately, polarization survey feedback could be used as a reward signal for reinforcement learning-based recommendation algorithms. This powerful emerging approach has the potential to learn what actually depolarizes, and continuously adapt to changes. Optimizing for such a signal may have unintended harmful consequences, so such a system would need to be continuously monitored in other ways, such as qualitative studies. In any case it may prove necessary to incorporate polarization measures into recommender systems to prevent the creation of conflict as a side effect of optimization (D'Amour et al., 2020).

It is unknown whether this sort of feedback-driven intervention would succeed in reducing the average dislike of the outgroup as compared to doing nothing, or more broadly whether intervening in platform recommenders will be an effective depolarization strategy within the complex and dynamic media ecosystem of any particular community, but there is reason to suspect this is possible. At the very least, the collection of individual-level affective polarization survey data provides a managerial incentive in the direction of depolarization. Nonetheless, the use of affective polarization survey data to drive platform recommender systems is a theoretically grounded, technically feasible, and potentially robust strategy for a social media depolarization intervention which deserves further study.

## Acknowledgements

This work was first presented at the BRaVE project's Exploring Societal Resilience to Online Polarization and Extremism workshop. The author thanks workshop participants and the anonymous reviewers for insightful feedback.

# References

Aggarwal, C. C. (2016). *Recommender Systems*. Springer. https://doi.org/10.1007/978-3-319-29659-3

Allcott, H., Braghieri, L., Eichmeyer, S., & Gentzkow, M. (2020). The welfare effects of social media. *American Economic Review*, *110*(3), 629–676. https://doi.org/10.1257/aer.20190658

American Political Science Association. (1950). Summary of Conclusions and Proposals. *American Political Science Review*, *44*(3), 1–14. https://www.jstor.org/stable/1950998

Angrist, J. D., & Pischke, J.-S. (2009). *Mostly Harmless Econometrics: An Empiricist's Companion*. Princeton University Press.

Asimovic, N., Nagler, J., Bonneau, R., & Tucker, J. A. (2021). *Testing the effects of Facebook usage in an ethnically polarized setting*. *118*(25). https://doi.org/10.1073/pnas.2022819118/-/DCSupplemental.y

Bail, C. A., Argyle, L. P., Brown, T. W., Bumpus, J. P., Chen, H., Hunzaker, M. B. F., Mann, M., Lee, J., Volfovsky, A., & Merhout, F. (2018). Exposure to opposing views on social media can increase political polarization. *PNAS*, *115*(37), 9216–9221. https://doi.org/https://doi.org/10.1073/pnas.1804840115

Baugut, P., & Neumann, K. (2020). Online propaganda use during Islamist radicalization. *Information Communication and Society*, *23*(11), 1570–1592. https://doi.org/10.1080/1369118X.2019.1594333

Berry, G., & Taylor, S. J. (2017). Discussion quality diffuses in the digital public square. *26th International World Wide Web Conference, WWW 2017*, 1371–1380. https://doi.org/10.1145/3038912.3052666

Boxell, L., Gentzkow, Matthew, & Shapiro, J. M. (2020). Cross-Country Trends in Affective Polarization. In *NBER Working Paper* (Issue June). https://www.nber.org/papers/w26669

Boxell, L., Gentzkow, M., & Shapiro, J. (2017). *Is the Internet Causing Political Polarization? Evidence from Demographics*. https://doi.org/10.3386/w23258

Bozdag, E., & van den Hoven, J. (2015). Breaking the filter bubble: democracy and design. *Ethics and Information Technology*, *17*(4), 249–265. https://doi.org/10.1007/s10676-015-9380-y

Bruns, A. (2019). *Are Filter Bubbles Real?* Polity.

Build Up. (2019). *The Commons: an intervention to depolarize political conversations on Twitter and Facebook in the USA*. https://howtobuildup.org/wp-content/uploads/2020/04/TheCommons-2019-Report_final.pdf

Castells, P., Hurley, N. J., & Vargas, S. (2015). Novelty and diversity in recommender systems. In *Recommender Systems Handbook, Second Edition* (pp. 881–918). Springer US. https://doi.org/10.1007/978-1-4899-7637-6_26

Celis, L. E., Kapoor, S., Salehi, F., & Vishnoi, N. (2019). Controlling Polarization in Personalization. *FAT\* '19: Conference on Fairness, Accountability, and Transparency*, 160–169. https://doi.org/10.1145/3287560.3287601

Collins, R. (2012). C-escalation and D-escalation: A theory of the time-dynamics of conflict. *American Sociological Review*, *77*(1), 1–20. https://doi.org/10.1177/0003122411428221

Creemers, R. (2017). Cyber China: Upgrading propaganda, public opinion work and social management for the twenty-first century. *Journal of Contemporary China*, *26*(103), 85–100. https://doi.org/10.1080/10670564.2016.1206281

D'Amour, A., Heller, K., Moldovan, D., Adlam, B., Alipanahi, B., Beutel, A., Chen, C., Deaton, J., Eisenstein, J., Hoffman, M. D., Hormozdiari, F., Houlsby, N., Hou, S., Jerfel, G., Karthikesalingam, A., Lucic, M., Ma, Y., McLean, C., Mincu, D., … Sculley, D. (2020). *Underspecification Presents Challenges for Credibility in Modern Machine Learning*. http://arxiv.org/abs/2011.03395

Deutsch, M. (1969). Conflicts: Productive and Destructive. *Journal of Social Issues*, *25*(1), 7–42. https://doi.org/10.1111/j.1540-4560.1969.tb02576.x

Diamond, L., Drutman, L., Lindberg, T., Kalmoe, N. P., & Mason, L. (2020). Americans Increasingly Believe Violence is Justified if the Other Side Wins. *Politico*. https://www.politico.com/news/magazine/2020/10/01/political-violence-424157

Draca, M., & Schwarz, C. (2018). How Polarized are Citizens? Measuring Ideology from the Ground-Up. *SSRN Electronic Journal*. https://doi.org/10.2139/ssrn.3154431

Druckman, J. N., Gubitz, S. R., Levendusky, M. S., & Lloyd, A. M. (2019). How incivility on partisan media (De)polarizes the electorate. *Journal of Politics*, *81*(1), 291–295. https://doi.org/10.1086/699912

Druckman, J. N., & Levendusky, M. S. (2019). What do we measure when we measure affective polarization? *Public Opinion Quarterly*, *83*(1), 114–122. https://doi.org/10.1093/poq/nfz003

Duarte, N., Llanso, E., & Loup, A. (2017). *Mixed Messages? The Limits of Automated Social Media Content Analysis*. Center for Democracy and Technology. https://cdt.org/insights/mixed-messages-the-limits-of-automated-social-media-content-analysis/

Feezell, J. T., Wagner, J. K., & Conroy, M. (2021). Exploring the effects of algorithm-driven news sources on political behavior and polarization. *Computers in Human Behavior*, *116*(November 2020), 106626. https://doi.org/10.1016/j.chb.2020.106626

Finkel, E. J., Bail, C. A., Cikara, M., Ditto, P. H., Iyengar, S., Klar, S., Mason, L., McGrath, M. C., Nyhan, B., Rand, D. G., Skitka, L. J., Tucker, J. A., Van Bavel, J. J., Wang, C. S., & Druckman, J. N. (2020). Political sectarianism in America. *Science*, *370*(6516), 533–536. https://doi.org/10.1126/science.abe1715

Fixdal, M. (2012). *Just Peace: How Wars Should End*. Palgrave Macmillan. https://doi.org/10.4324/9781351155762-10

Fletcher, R., & Nielsen, R. K. (2018). Are people incidentally exposed to news on social media? A comparative analysis. *New Media and Society*, *20*(7), 2450–2468. https://doi.org/10.1177/1461444817724170

Gentzkow, M., Shapiro, J., & Taddy, M. (2017). *Measuring Polarization in High-Dimensional Data: Method and Application to Congressional Speech* (No. 22423). http://www.nber.org/data-appendix/w22423

Griffioen, N., van Rooij, M., Lichtwarck-Aschoff, A., & Granic, I. (2020). Toward improved methods in social media research. *Technology, Mind, and Behavior*, *1*(1). https://doi.org/10.1037/tmb0000005

Guess, A., Lyons, B., Nyhan, B., & Reifler, J. (2018). Avoiding the Echo Chamber about Echo Chambers: Why selective exposure to like-minded political news is less prevalent than you think. *Knight Foundation*. https://kf-site-production.s3.amazonaws.com/media_elements/files/000/000/133/original/Topos_KF_White-Paper_Nyhan_V1.pdf

Halevy, A., Ferrer, C. C., Ma, H., Ozertem, U., Pantel, P., Saeidi, M., Silvestri, F., & Stoyanov, V. (2020). *Preserving Integrity in Online Social Networks*. http://arxiv.org/abs/2009.10311

Hare, C., & Poole, K. T. (2014). The polarization of contemporary American politics. *Polity*, *46*(3), 411–429. https://doi.org/10.1057/pol.2014.10

Hautakangas, M., & Ahva, L. (2018). Introducing a New Form of Socially Responsible Journalism: Experiences from the Conciliatory Journalism Project. *Journalism Practice*, *12*(6), 730–746. https://doi.org/10.1080/17512786.2018.1470473

Helberger, N. (2019). On the Democratic Role of News Recommenders. *Digital Journalism*, *7*(8), 993–1012. https://doi.org/10.1080/21670811.2019.1623700

Helberger, N., Karppinen, K., & D'Acunto, L. (2018). Exposure diversity as a design principle for recommender systems. *Information Communication and Society*, *21*(2), 191–207. https://doi.org/10.1080/1369118X.2016.1271900

Hofstetter, J.-S. (2021). *Digital Technologies , Peacebuilding and Civil Society* (No. 114). https://ict4peace.org/activities/digital-technologies-peacebuilding-and-civil-society-by-julia-hofstetter-senior-advisor-ict4peace/

Hosseinmardi, H., Ghasemian, A., Clauset, A., Rothschild, D. M., Mobius, M., & Watts, D. J. (2020). *Evaluating the scale, growth, and origins of right-wing echo chambers on YouTube*. http://arxiv.org/abs/2011.12843

Ie, E., Hsu, C., Mladenov, M., Jain, V., Narvekar, S., Wang, J., Wu, R., & Boutilier, C. (2019). *RecSim: A Configurable Simulation Platform for Recommender Systems*. http://arxiv.org/abs/1909.04847

Iyengar, S., Lelkes, Y., Levendusky, M., Malhotra, N., & Westwood, S. J. (2018). The Origins and Consequences of Affective Polarization in the United States. *Annual Review of Political Science*, 1–35. https://doi.org/10.1146/annurev-polisci-051117-073034

Iyengar, S., & Westwood, S. J. (2015). Fear and Loathing across Party Lines: New Evidence on Group Polarization. *American Journal of Political Science*, *59*(3), 690–707. https://doi.org/10.1111/ajps.12152

Jackson, A. (2005). Falling from a great height: Principles of good practice in performance measurement and the perils of top down determination of performance indicators. *Local Government Studies*, *31*(1), 21–38. https://doi.org/10.1080/0300393042000332837

Jacobs, A. Z., & Wallach, H. (2019). *Measurement and Fairness*. http://arxiv.org/abs/1912.05511

Jeong, H. W. (2019). Conflict Transformation. In S. Byrne, T. Matyók, I. M. Scott, & J. Senehi (Eds.), *Routledge Companion to Peace and Conflict Studies* (pp. 25–34). Routledge. https://doi.org/10.4324/9781315182070

Jhaver, S., Vora, P., & Bruckman, A. (2017). Designing for Civil Conversations: Lessons Learned from ChangeMyView. *GVU Technical Report*. https://smartech.gatech.edu/handle/1853/59080

Jiang, R., Chiappa, S., Lattimore, T., György, A., & Kohli, P. (2019). *Degenerate Feedback Loops in Recommender Systems*. https://doi.org/10.1145/3306618.3314288

Keller, D. (2018). Internet platforms: Observations on speech, danger, and money. *Hoover Institution, Aegis Series Paper No. 1807*, 5–8. https://lawfareblog.com/internet-platforms-observations-speech-danger-and-money

Kiley, J. (2017). *In polarized era, fewer Americans hold a mix of conservative and liberal views*. Pew Research Center. https://www.pewresearch.org/fact-tank/2017/10/23/in-polarized-era-fewer-americans-hold-a-mix-of-conservative-and-liberal-views/

Kim, Y. (2015). Does disagreement mitigate polarization? How selective exposure and disagreement affect political polarization. *Journalism and Mass Communication Quarterly*, *92*(4), 915–937. https://doi.org/10.1177/1077699015596328

Kim, Y., & Kim, Y. (2019). Incivility on facebook and political polarization: The mediating role of seeking further comments and negative emotion. *Computers in Human Behavior*, *99*(February), 219–227. https://doi.org/10.1016/j.chb.2019.05.022

King, D. S., & Smith, R. M. (2008). *Strange Bedfellows? Polarized Politics? The Quest for Racial Equity in Contemporary America*. September, 686–703. https://doi.org/10.1177/1065912908322410

King, G., Pan, J., & Roberts, M. E. (2017). How the Chinese government fabricates social media posts for strategic distraction, not engaged argument. *American Political Science Review*, *111*(3), 484–501. https://doi.org/10.1017/S0003055417000144

Krueger, D. S., Maharaj, T., & Leike, J. (2020). *Hidden incentives for auto-induced distributional shift*. https://arxiv.org/abs/2009.09153

Layman, G. C., Carsey, T. M., Green, J. C., Herrera, R., & Cooperman, R. (2010). Activists and conflict extension in American party politics. *American Political Science Review*, *104*(2), 324–346. https://doi.org/10.1017/S000305541000016X

Lederach, J. P. (2014). *The Little Book of Conflict Transformation*.

Ledwich, M., & Zaitsev, A. (2019). Algorithmic Extremism: Examining YouTube's Rabbit Hole of Radicalization. *First Monday*. https://doi.org/10.5210/fm.v25i3.10419

Lee, A. H. Y. (2020). How the Politicization of Everyday Activities Affects the Public Sphere: The Effects of Partisan Stereotypes on Cross-Cutting Interactions. *Political Communication*, *00*(00), 1–20. https://doi.org/10.1080/10584609.2020.1799124

Lee, F. E. (2015). How party polarization affects governance. *Annual Review of Political Science*, *18*, 261–282. https://doi.org/10.1146/annurev-polisci-072012-113747

Lelkes, Y., Sood, G., & Iyengar, S. (2017). The Hostile Audience: The Effect of Access to Broadband Internet on Partisan

Affect. *American Journal of Political Science*, *61*(1), 5–20. https://doi.org/10.1111/ajps.12237
Levy, R. (2020). Social Media, News Consumption, and Polarization: Evidence from a Field Experiment. *SSRN Electronic Journal*. https://doi.org/10.2139/ssrn.3653388
Li, R. Z., Urbano, J., & Hanjalic, A. (2021). *Leave No User Behind: Towards Improving the Utility of Recommender Systems for Non-mainstream Users*. https://doi.org/10.1145/3437963.3441769 10.1145/3437963.3441769 10.1145/3437963.3441769
Litt, E., Zhao, S., Kraut, R., & Burke, M. (2020). What Are Meaningful Social Interactions in Today's Media Landscape? A Cross-Cultural Survey. *Social Media + Society*, *6*(3), 205630512094288. https://doi.org/10.1177/2056305120942888
Loecherbach, F., Moeller, J., Trilling, D., & van Atteveldt, W. (2020). The Unified Framework of Media Diversity: A Systematic Literature Review. *Digital Journalism*, *8*(5), 605–642. https://doi.org/10.1080/21670811.2020.1764374
Manheim, D., & Garrabrant, S. (2018). *Categorizing Variants of Goodhart's Law*. 1–10. http://arxiv.org/abs/1803.04585
Martherus, J. L., Martinez, A. G., Piff, P. K., & Theodoridis, A. G. (2021). Party Animals? Extreme Partisan Polarization and Dehumanization. *Political Behavior*, *43*(2), 517–540. https://doi.org/10.1007/s11109-019-09559-4
Martin, G. J., & Yurukoglu, A. (2016). Bias in Cable News: Persuasion and Polarization. In *NBER Working Papers* (No. 20798). http://www.nber.org/papers/w20798
Matias, J. N. (2019). Preventing harassment and increasing group participation through social norms in 2,190 online science discussions. *Proceedings of the National Academy of Sciences of the United States of America*, *116*(20), 9785–9789. https://doi.org/10.1073/pnas.1813486116
McCoy, J., Rahman, T., & Somer, M. (2018). Polarization and the Global Crisis of Democracy: Common Patterns, Dynamics, and Pernicious Consequences for Democratic Polities. *American Behavioral Scientist*, *62*(1), 16–42. https://doi.org/10.1177/0002764218759576
Melki, M., & Pickering, A. (2020). Polarization and corruption in America. *European Economic Review*, *124*, 103397. https://doi.org/10.1016/j.euroecorev.2020.103397
Melki, M., & Sekeris, P. G. (2019). Media-driven polarization. Evidence from the US. *Economics*, *13*, 0–14. https://doi.org/10.5018/economics-ejournal.ja.2019-34
Mladenov, M., Meshi, O., Ooi, J., Schuurmans, D., & Boutilier, C. (2019). Advantage Amplification in Slowly Evolving Latent-State Environments. *IJCAI International Joint Conference on Artificial Intelligence*, *2019-Augus*, 3165–3172. https://doi.org/10.24963/ijcai.2019/439
Möller, J., Trilling, D., Helberger, N., & van Es, B. (2018). Do not blame it on the algorithm: an empirical assessment of multiple recommender systems and their impact on content diversity. *Information, Communication & Society*, *21*(7), 959–977. https://doi.org/10.1080/1369118X.2018.1444076
Mouffe, C. (2002). Which Public Sphere for a Democratic Society? *Theoria: A Journal of Social and Political Theory*, *22*(99), 55–65. https://www.jstor.org/stable/41802189
Munger, K., & Phillips, J. (2020). Right-Wing YouTube: A Supply and Demand Perspective. *International Journal of Press/Politics*. https://doi.org/10.1177/1940161220964767
Noever, D. (2018). *Machine Learning Suites for Online Toxicity Detection*. https://arxiv.org/abs/1810.01869
Noorshams, N., Verma, S., & Hofleitner, A. (2020). TIES: Temporal Interaction Embeddings for Enhancing Social Media Integrity at Facebook. *Proceedings of the ACM SIGKDD International Conference on Knowledge Discovery and Data Mining*, 3128–3135. https://doi.org/10.1145/3394486.3403364
Paolini, S., Harwood, J., & Rubin, M. (2010). Negative intergroup contact makes group memberships salient: Explaining why intergroup conflict endures. *Personality and Social Psychology Bulletin*, *36*(12), 1723–1738. https://doi.org/10.1177/0146167210388667
Pettigrew, T. F., & Tropp, L. R. (2006). A meta-analytic test of intergroup contact theory. *Journal of Personality and Social Psychology*, *90*(5), 751–783. https://doi.org/10.1037/0022-3514.90.5.751
Pew Research Center. (2014). *Political Polarization in the American Public*. https://www.pewresearch.org/politics/2014/06/12/political-polarization-in-the-american-public/
Prior, M. (2013). Media and Political Polarization. *Annu. Rev. Polit. Sci*, *16*, 101–127. https://doi.org/10.1146/annurev-polisci-100711-135242
Prior, M., & Stroud, N. J. (2015). Using Mobilization, Media, and Motivation to Curb Political Polarization. In N. Persily (Ed.), *Solutions to Political Polarization in America*. Cambridge University Press. https://doi.org/https://doi.org/10.1017/CBO9781316091906.013
Ramsbotham, O., Tom Woodhouse, & Hugh Miall. (2016). *Contemporary Conflict Resolution, 4th Edition*. Wiley.
Ribeiro, M. H., Ottoni, R., West, R., Almeida, V. A. F., & Wagner Meira, W. M. (2020). Auditing radicalization pathways on YouTube. *FAT\* 2020 - Proceedings of the 2020 Conference on Fairness, Accountability, and Transparency*, 131–141. https://doi.org/10.1145/3351095.3372879
Ripley, A. (2018). *Complicating the Narratives*. Solutions Journalism Network. https://thewholestory.solutionsjournalism.org/complicating-the-narratives-b91ea06ddf63
Ripley, A. (2021). *High Conflict: Why We Get Trapped and How We Get Out*. Simon & Schuster.
Rychwalska, A., & Roszczyńska-Kurasińska, M. (2018). Polarization on Social Media: When Group Dynamics Leads to Societal Divides. *Proceedings of the 51st Hawaii International Conference on System Sciences*. https://doi.org/10.24251/hicss.2018.263
Schirch, L. (2020). *Social Media Impacts on Conflict Dynamics: A Synthesis of Ten Case Studies & a Peacebuilding Plan for Tech* (No. 73; Toda Peace Institute). https://toda.org/policy-briefs-and-resources/policy-briefs/social-media-impacts-on-conflict-dynamics-a-synthesis-of-ten-case-studies-and-a-peacebuilding-plan-for-tech.html
Somer, M., & McCoy, J. (2019). Transformations through Polarizations and Global Threats to Democracy. *The Annals of the American Academy of Political and Social Science*, *681*(1), 8–22. https://doi.org/10.1177/0002716218818058

Stoica, A. A., & Chaintreau, A. (2019). Hegemony in social media and the effect of recommendations. *The Web Conference 2019 - Companion of the World Wide Web Conference, WWW 2019*, *2*, 575–580. https://doi.org/10.1145/3308560.3317589

Stray, J. (2020). Aligning AI Optimization to Community Well-being. *International Journal of Community Well-Being*. https://doi.org/10.1007/s42413-020-00086-3

Stray, J. (2021). *Beyond Engagement: Aligning Algorithmic Recommendations With Prosocial Goals*. Partnership on AI. https://www.partnershiponai.org/beyond-engagement-aligning-algorithmic-recommendations-with-prosocial-goals/

Stray, J., Adler, S., & Hadfield-Menell, D. (2020). What are you optimizing for ? Aligning Recommender Systems with Human Values. *Participatory Approaches to Machine Learning Workshop, ICML 2020*. https://participatoryml.github.io/papers/2020/42.pdf

Stroud, N. J., Muddiman, A., & Scacco, J. M. (2017). Like, recommend, or respect? Altering political behavior in news comment sections. *New Media and Society*, *19*(11), 1727–1743. https://doi.org/10.1177/1461444816642420

Suhay, E., Bello-Pardo, E., & Maurer, B. (2018). The Polarizing Effects of Online Partisan Criticism: Evidence from Two Experiments. *International Journal of Press/Politics*, *23*(1), 95–115. https://doi.org/10.1177/1940161217740697

Taber, C. S., & Lodge, M. (2006). Motivated skepticism in the evaluation of political beliefs (2006). *American Journal of Political Science*, *50*(3), 755–769. https://doi.org/10.1111/j.1540-5907.2006.00214.x

Tellidis, I., & Kappler, S. (2016). Information and communication technologies in peacebuilding: Implications, opportunities and challenges. *Cooperation and Conflict*, *51*(1), 75–93. https://doi.org/10.1177/0010836715603752

Thomas, R. L., & Uminsky, D. (2020). Reliance on Metrics is a Fundamental Challenge for AI. *Ethics of Data Science Conference*. https://arxiv.org/abs/2002.08512

van Stekelenburg, J. (2014). Going all the way: Politicizing, polarizing, and radicalizing identity offline and online. *Sociology Compass*, *8*(5), 540–555. https://doi.org/10.1111/soc4.12157

Wagner, K. (2019, March). Inside Twitter's ambitious plan to change the way we tweet. *Vox*. https://www.vox.com/2019/3/8/18245536/exclusive-twitter-healthy-conversations-dunking-research-product-incentives

Whitesides, J. (2017, February 7). From disputes to a breakup: wounds still raw after U.S. election. *Reuters*. https://www.reuters.com/article/us-usa-trump-relationships-insight/from-disputes-to-a-breakup-wounds-still-raw-after-u-s-election-idUSKBN15M13L

York, J., & Zuckerman, E. (2019). Moderating the Public Sphere. In R. F. Jørgensen (Ed.), *Human Rights in the Age of Platforms* (pp. 137–161). MIT Press.

Zhao, Z., Hong, L., Wei, L., Chen, J., Nath, A., Andrews, S., Kumthekar, A., Sathiamoorthy, M., Yi, X., & Chi, E. (2019). Recommending what video to watch next. *Proceedings of the 13th ACM Conference on Recommender Systems*, 43–51. https://doi.org/10.1145/3298689.3346997

Zuiderveen Borgesius, F. J., Trilling, D., Möller, J., Bodó, B., de Vreese, C. H., & Helberger, N. (2016). Should we worry about filter bubbles? *Internet Policy Review*, *5*(1), 1–16. https://doi.org/10.14763/2016.1.401

Zuiderveen Borgesius, F. J., Trilling, D., Möller, J., Bodó, B., De Vreese, C. H., & Helberger, N. (2016). Should we worry about filter bubbles? *Internet Policy Review*, *5*(1), 1–16. https://doi.org/10.14763/2016.1.401